\documentclass[useAMS,usenatbib]{mn2e}

\usepackage{graphicx}
\usepackage{amsmath}
\usepackage{amsfonts}
\usepackage{amssymb}
\usepackage{rotating}
\usepackage[english]{babel}

\def \kms{\rm{km}$\rm{s}^{-1}$}

\def \cm{~\rm{cm}}
\def \s{~\rm{s}}
\def \km{~\rm{km}}
\def \kms{~\rm{km}~{\rm s}^{-1}}

\def \K{~\rm{K}}

\def \AU{~\rm{AU}}
\def \erg{~\rm{erg}}

\def \yr{~\rm{yr}}

\def\actaa{Acta Astron.}%
\def\apj{ApJ}%
\def\apjl{ApJ}%
\def\apjs{ApJS}%
\def\aap{A\&A}%
\def\mnras{MNRAS}%
\def\na{New A}%
\def\nat{Nature}%
\def \actaa{Acta Astron.}

\title{Explaining the type Ia supernova PTF 11kx with a {{{{  {violent-prompt merger} }}}} scenario}

\author[N. Soker et al.]{Noam Soker,$^{1}$\thanks{soker@physics.technion.ac.il}
                         Amit Kashi,$^{2}$
                         Enrique Garc\'\i a--Berro,$^{3,4}$
                         Santiago Torres$^{3,4}$ and
                         \newauthor
                         Judit Camacho$^{3,4}$\\
                     $^1$Department of Physics, Technion ---
                         Israel Institute of Technology,
                         Haifa 32000, Israel\\
                     $^2$Department of Physics and Astronomy,
                         University of Nevada,
                         Las Vegas, 4505 S. Maryland Pkwy,
                         Las Vegas, NV, 89154-4002, USA\\
                     $^3$Departament de F\'\i sica Aplicada,
                         Universitat Polit\`ecnica de Catalunya,
                         c/Esteve Terrades 5,
                         08860 Castelldefels, Spain\\
                     $^4$Institute for Space Studies of Catalonia,
                         c/Gran Capit\`a 2--4,
                         Edif. Nexus 104,
                         08034 Barcelona, Spain}

\begin{document}

\date{\today}

\maketitle

\begin{abstract}
We argue that the multiple  shells of circumstellar material (CSM) and
the supernovae (SN)  ejecta interaction with the CSM  starting 59 days
after  the explosion  of the  Type Ia  SN (SN  Ia) PTF~11kx,  are best
described by a violent prompt merger.
In this prompt merger scenario the common envelope (CE) phase is terminated by a merger of  a WD
companion with the hot core of a massive asymptotic giant (AGB) star.
In most cases the WD is {{ disrupted}} and accreted onto the more massive core.
However, in the rare cases where the merger takes place when the WD is  denser than  the  core, the  core will  be {{ disrupted}}  and
accreted onto the cooler WD.   In such cases the explosion might occur
with no appreciable delay, i.e., months to years after the termination of the
CE phase.   This,  we  propose,  might be  the evolutionary route  that could lead  to the explosion  of PTF~11kx.
This scenario can account  for the very massive CSM  within $\sim 1000 \AU$
of the exploding PTF~11kx star,  for the presence of hydrogen, and for
the presence of shells in the CSM.
\end{abstract}


\section{INTRODUCTION}
\label{sec:intro}

Observations and  theoretical studies cannot teach us  yet whether all
three scenarios  for the  formation of Type  Ia supernova (SN  Ia), or
only  one  or  two  of  them  can  work  ---  e.g.,  \cite{Livio2001},
\cite{Maoz2010},  \cite{Howell2011}.   These  three basic  theoretical
scenarios can be described as follows.  ($i$) In the single-degenerate
(SD)   scenario  (e.g.,   \citealt{Whelan1973};  \citealt{Nomoto1982};
\citealt{Han2004})  a  WD  grows  in  mass through  accretion  from  a
non-degenerate stellar  companion.  However, the mass  increase of the
WD seems to be very limited (e.g., \citealt{Idan2012}).  ($ii$) In the
double-degenerate (DD)  scenario (\citealt{Webbink1984, Iben1984}; see
\citealt{vanKerkwijk2010}  for  a   paper  on  sub-Chandrasekhar  mass
remnants)  two WDs  merge  after losing  energy  and angular  momentum
through  the  radiation  of gravitational  waves  \citep{Tutukov1979}.
Some  list  the  ``double-detonation''  mechanism  \citep{Woosley1994,
Livne1995}  as  a separate  channel,  although  it  involves two  WDs.
($iii$) In the  core-degenerate (CD) scenario for the  formation of SN
Ia the Chandrasekhar ($M_{\rm  Ch}$) or super-Chandrasekhar mass WD is
formed at  the termination  of the CE  phase, or during  the planetary
nebula phase, from a  merger of a WD companion with the  hot core of a
massive  AGB star \citep{KashiSoker2011,  IlkovSoker2012a, Soker2011}.
The merger  of a WD with  the core of an  AGB star was  studied in the
past  (\citealt{Sparks1974, Livio2003,  Tout2008}).  \citet{Livio2003}
suggested that the merger of the WD with the AGB core leads to a SN Ia
that  occurs at the  end of  the CE  phase or  shortly after,  and can
explain the  presence of  hydrogen lines.  {{ However,}} due  to its  rapid rotation
(e.g.,           \citealt{Anand1965};          \citealt{Ostriker1968};
\citealt{Uenishi2003};  \citealt{Yoon2005};  \citealt{Lorenetal2009}),
and   possibly   very    strong   central   magnetic   fields   (e.g.,
\citealt{Kundu2012}),  the explosion of  a WD  with $M\ge  M_{\rm Ch}$
might be substantially delayed.

The  recently observed  SN Ia  PTF~11kx \citep{Dilday2012}  has narrow
lines, including  hydrogen lines, and indications  of interaction with
a  massive circumstellar medium  (CSM) which starts 59  days after
the  explosion.    \cite{Dilday2012}  argued  that   PTF~11kx  can  be
explained by  the SD scenario for  SN Ia.  They dismiss  the merger {{ scenario}} as
suggested by \citet{Livio2003} on several grounds. {{ In particular,} they claim it
cannot account  for several CSM  shells.  We find this  unjustified in
section \ref{sec:CSM}.
We further discuss the {{ properties}} of the CSM in section \ref{sec:radiated}, where we calculate
the total radiated energy {{ arising}} from the collision of the exploding gas with the CSM.

The destruction  of the WD and  its accretion onto the  core while the
core  is  still  hot  might   prevent  an  early  ignition  of  carbon
\citep{Yoon2007}, which is  one of the theoretical problems  of the DD
scenario  (e.g., \citealt{SaioNomoto2004}).  However,  in the  case of
PTF~11kx  ignition of  the  merger product  is  required quite  early,
within $\sim 30  \yr$ from the CE ejection.   We therefore consider in
section \ref{sec:massive} the possibility that in the case of PTF~11kx
the core was destructed onto  the cooler WD.  In {{ those}} cases where {{ an} explosion
occurs shortly after merger in the  CE, there is no delay by spin-down
or emission  of gravitational  waves.
{{{{  { This is called a violent-prompt merger, which might be considered as a sub-version of the DD and CD scenarios.} }}}}
In  section \ref{sec:ignition} we show that  {{ in}} these cases
 a massive envelope {{ can be ejected}}, and {{ that this scenario}} can account  for the frequency of such SN Ia.
Our summary is in section \ref{sec:summary}.

\section{A CASE FOR A MASSIVE CIRCUMSTELLAR MEDIUM (CSM)}
\label{sec:CSM}

Starting 59  days after  the explosion of  PTF~11kx, \cite{Dilday2012}
detected a group  of absorption lines (Na  I, Fe II, Ti II,  and He I)
and H$\alpha$  and H$\beta$ P-cygni lines  at a velocity  shift of $65
\kms$.  The Ca II H\&K absorption  line was detected at an outflow velocity
of $100 \kms$  and a group of  emission lines (H, He, Fe,  Ti) at $100
\kms$.  This,  according to \cite{Dilday2012},  indicates the presence
of multiple  CSM shells  interacting with the  SN ejecta.  Adopting an
ejecta  velocity  of  $25\,000  \km  \s^{-1}$ they  deduced  that  the
ejecta-CSM interaction starts at a distance of $\sim 10^{16} \cm$.

\cite{Dilday2012} suggest that PTF~11kx is  a type Ia SN formed by the
SD scenario.  A claim that was  raised also in the theoretical work of
\cite{Hachisu2012}.   In  the scenario  of  \cite{Dilday2012} the  CSM
structure is assumed to originate from recent recurrent nova eruptions
whose  ejecta sweep  the  wind of  a  giant companion.   Based on  the
saturated Ca~II~HK  absorption lines \cite{Dilday2012}  calculated the
CSM mass to  be $M_{\rm CSM} \simeq 5.3 k \,{\rm M_{\sun}}$,  where $k$ is the
covering  fraction.  For  the absence  of an  interaction with  a very
massive CSM,  they argue that $k  \ll 1$.  The relevant  radius of the
photosphere at  $t=20$~d is $r_{\rm  ph} = 1.6 \times  10^{15} \cm$.
The  location of  the absorbing  gas  according to  their analysis  is
$r_{\rm Ca} = 1.3 \times 10^{16} \cm$.  Taking the absorbing gas to be
in a  torus (as  they do)  of larger radius  $r_{\rm Ca}$  and smaller
radius  $r_{\rm ph}$, the  covering fraction  is $k  \ga r_{\rm  ph} /
r_{\rm  Ca}  =   0.12$.   We  conclude,  based  on   the  analysis  of
\cite{Dilday2012}, that $M_{\rm CSM} > 0.6 \, {\rm M_{\sun}}$.

We  can also  estimate $M_{\rm  CSM}$ from  the  H$\alpha$ luminosity.
\cite{Dilday2012}  list  $L_{{\rm  H}\alpha}$ at  5 epochs,  with a
maximum of  $L_{{\rm H}\alpha} (88\,  {\rm d})=1.84 \times  10^{40} \erg
\s^{-1}$.  We  take a case  B recombination for solar  composition gas
and derive
\begin{equation}
M_{{\rm H}\alpha} \simeq 0.2 \left( \frac {V}{10^{48} \cm^3} \right)^{1/2} {\rm M_{\sun}},
\label{eq:massh}
\end{equation}
where $V$  is a volume, and  where we assumed a  constant density. The
volume  was  scaled  according   to  the  torus  discussed  above,  If
ionization is not complete then the mass {{ could be even larger}}. This estimate
does not  depend on  the calcium abundance.   A value of  $M_{\rm CSM}
\simeq M_{{\rm H}\alpha} >0.2 {\rm M_{\sun}}$ at a distance of $\sim 10^{16}
\cm$ is larger than what most SD scenarios predict.

Our motivation to explore the {{{{  {violent-prompt merger} }}}}  scenario to account for a massive CSM
comes from  other SN  Ia as well.   SN~2002ic \citep{Hamuy2003}  had a
strong and broad, FWHM of  $> 1000 \kms$, H$\alpha$ emission resulting
from SN  ejecta-CSM interaction,  with CSM mass  of $\sim  6 {\rm M_{\sun}}$
(\citealt{Wang2004}).   \cite{Nomoto2005,  Nomoto2004} suggested  that
the lightcurve can be explained by a non-spherical $\sim 1.3 {\rm M_{\sun}}$
CSM.   \cite{Livio2003}  proposed  that  SN~2002ic represents  a  rare
subtype of  the DD scenario in  which the explosion  occurs by core-WD
merger immediately  following the CE  phase.


SN~2005gj    is    another    SN    Ia    with    hydrogen    emission
(\citealt{Aldering2006},  \citealt{Prieto2007}), and an  aspherical or
clumpy  CSM  \citep{Aldering2006}  with  an  estimated  mass  of  $\ga
0.016-0.16 {\rm M_{\sun}}$ \citep{Aldering2006}.  More emission lines of
hydrogen  and   helium  were  detected  with   time  since  explosion,
indicating the presence of a few CSM layers.

The multi-layer  aspherical CSM in the  above SNe can  be explained in
the frame of the {{{{  {violent-prompt merger } }}}}  scenario.  (1) Before the CE { phase} there is the ordinary
mass loss  of the  AGB star, but  enhanced by tidal  interaction.  (2)
During  the CE  ejection a  spiral structure  can form  several shells
along the  line of sight, and  in particular in  the equatorial plane.
Some   numerical    simulations   ---   e.g.,   \cite{RickerTaam2012};
\cite{Sandquist1998} ---  show that  the CE mass  loss occurs  in more
than one episode,  i.e., there are a few peaks in  the mass loss rate.
In particular, along each direction there can be several shells due to
the spiral structure that is formed  during the CE phase, and that can
later have imprint in the CSM.   (3) The merger process can lead to an
equatorial mass  loss event that  ejects $\sim 0.1$--$0.5\,  {\rm M_{\sun}}$.  It
can accelerate some previously ejected gas to higher velocities, e.g.,
$\sim 100 \kms$.   (4) As observed in some  post-CE planetary nebulae,
e.g., Sp$-$1 and NGC 2346, the mass close to the star is concentrated in
the equatorial plane.  The rest  of the mass forms a bipolar structure
at much  larger distances.  In our  model the explosion  took place at
the center  of a pre-planetary nebulae.  Our  model therefore predicts
that further  interaction of the ejecta  with the CSM  will take place
over  the coming  years  to tens  of  years, and  that  the amount  of
accumulated mass in the CSM will be on the order of a solar mass.

\section{RADIATED ENERGY}
\label{sec:radiated}

We  calculate  the  excess  energy   of  PTF  11kx  ---  according  to
\cite{Dilday2012}   ---  over   the   ``normal''  SN   Ia  SN   2002er
(\citealt{Jha2007}).    We  extract  the   R-band  light   curve  from
\cite{Dilday2012} and  integrate the luminosity as a  function of time
from day $-7$ to day 99.   For PTF 11kx we take the luminosity between
day 64 and day 99 to be constant, set to the value of day 64.  We find
that over the  integrated time  the energy  in the  R band  of SN
2002er is $\sim 9.1 \times 10^{48}  \erg$, and the total energy in the
R band of  PTF 11kx is $\sim 2.1 \times 10^{49}  \erg$. Thus, in about
100 days PTF 11kx radiated $\sim  1.2 \times 10^{49} \erg$ more than a
typical SN Ia.   As this is only the R band,  the actual excess energy
value  can be larger. {{{{{ On the other hand by assuming that from day 64 to 99 the R-band luminosity does not decrease
we somewhat overestimate the radiated energy. }}}}}
Over a  period of  200 days  we will  take the
excess energy  that we attribute to  the collision of  the ejecta with
the CSM to be $\sim 2 \times 10^{49} \erg$.

Let us  take the ejecta  flowing with a  velocity of $v_{\rm  ej} \sim
10^4 \km \s^{-1}$  to hit the CSM residing in  a torus.  In estimating
the  kinetic  energy  that  is  transferred to  radiation  one  should
consider the geometry.  The CSM is  in a torus rather than a spherical
shell, and the shocked ejecta flows around it and cools adiabatically.
The fraction of the post-shock thermal energy that will be radiated is
the ratio $\eta \sim \tau_{\rm f}/(\tau_{\rm c} +\tau_{\rm f})$, where
$\tau_{\rm c}$ is the radiative cooling time and
\begin{equation}
\begin{split}
\tau_{\rm f} &\simeq \frac{ r_{\rm ph} }{v_{\rm ej}} \\
&= 2 \times 10^6
 \left( \frac{r_{\rm ph}} {2 \times 10^{15} \cm} \right)
\left( \frac{v_{\rm ej}} {10^4 \km \s^{-1}} \right)^{-1} \s
\label{eq:tauc}
\end{split}
\end{equation}
is the flow time scale.

For a direct shock the post-shock temperature is $T_{\rm ps} \sim 3 \times 10^{8} \K$.  The post-shock  density of the  ejecta and the
CSM have a similar electron density of $n_{\rm e} \sim 10^9 \cm^{-3}$.
For this  temperature and  density the cooling  time for  an optically
thin gas  (the optical  depth over  the small radius  of the  torus is
$\sim 0.1$--$0.5$) is $\tau_{\rm c} \simeq 2 \times 10^6
\s$.  Thus, a  fraction of $\eta \sim 0.5$ of  the thermal energy will
be  radiated.  In  the collision  of the  ejecta with  the CSM  only a
fraction $\zeta  < 1$ of  the kinetic energy  of the colliding  gas is
transferred to thermal energy because a large area of the shock on the
torus is oblique and because the  CSM is accelerated and takes part of
the kinetic energy.   Over all, the fraction of  the kinetic energy of
the ejecta that  is radiated is $k \eta \zeta \sim  0.05 \zeta$. For a
kinetic energy  of $10^{51}  \erg$ this amounts  to $5  \times 10^{49}
\zeta \erg$.  The value of $\zeta$  depends on the geometry of the CSM
and how it  is being changed during the  interaction.  The interaction
should last for  a time of $t_{\rm i}  > 1.3 \times 10^{16} \cm  / {\it v}_{\rm
ej} = 150$ days.  Therefore, the  energy radiated within a time of 200
days can be  $\sim 2 \times 10^{49} \erg$.  Hence,  a value of kinetic
to  thermal  conversion  ratio  of  $\zeta=0.4$ can  account  for  the
observed extra luminosity.

\section{ENERGY RELEASED IN THE MERGER PROCESS}
\label{sec:massive}

\subsection{The migration phase}
\label{sec:migration}

There are {{ various}} phases in the merger process \citep{KashiSoker2011}.
The dynamical {{ CE}} phase ends when most of the envelope is ejected, and the binary system ends at an orbital separation
of $\sim 1-3 {\rm ~R_{\sun}}$. The gravitational energy released in this process ejects part of the envelope, and
 a circumbinary disk is formed by the bound material, including fall back material. {{ This material}} interacts gravitationally with the binary system.
{{ Typical timescales for building-up the disk range from weeks to years,  but these are gross estimates}} as the  process requires further study.
{{ In any case, during this phase}} the semi-major axis is reduced to $a \simeq 0.5-1 {\rm ~R_{\sun}}$, while the eccentricity increases up to $e \la 1$.
{{ Consequently,}} a merger can occur during this migration phase because of the substantial decrease in the periastron distance,
despite the modest decrease in the semi-major axis.
The merger lasts for several weeks up to a year.
If the circumbinary disk is lost before {{ the}} merger {{ occurs}} the final evolution {{ of the merger is driven by the emission of gravitational waves}} --- see Fig.~7 in \cite{KashiSoker2011}.
{{f We note that the timescale}} for this will  be shorter due to the high eccentricity.
In any case, the merger is set up by substantially increasing the eccentricity with a very moderate decrease of the semi-major axis.
This implies that during the migration phase the binary system loses an amount of energy which is {{ approximately}} equal to what it has been lost during the dynamical CE phase,
e.g., $\sim 10^{48} \erg$.
{{{{{ So the total energy that is estimated to be radiated during the migration phase is $\sim 10^{48} \erg$. }}}}}
{{{{  {Moreover, the high eccentricity and monotonic increase in eccentricity imply also that the merger is likely to be violent,
as required in our violent-prompt merger scenario. }}}}

{{ This evolutionary route is different of that proposed by \cite{Chugai2004} who}} assumed that the binary has a circular orbit. {{ Accordingly, they}} argued against the post-CE merger on the ground that the spiraling-in
to a distance where gravitational waves take over, $a \sim 0.03 {\rm ~R_{\sun}}$, will release much more energy than
is observed in SNe 2002ic and 1997cy.
However, the migration process {{ described above}} removes the objections of \cite{Chugai2004} because of the modest decrease in orbital energy {{ during it. Actually,}}
the modest amount of released gravitational energy {{ can be}} carried away by the mass ejected from the circumbinary disk and by radiation.
{{ Additionally,}} we note that this mass loss episode can form another circumbinary shell, with which the explosion {{ can}} interact later on.

There are two possible channels to ignite the WD-core merger product: ($i$) A violent merger where ignition occurs within
minutes \citep{Pakmoretal2011, Pakmoretal2012a, Pakmoretal2012b}, and ($ii$) merging and relaxing within years {{ up}} to the point {{ where}}
the WD {{ becomes}} unstable and {{ is ignited}}.
We first note that during the merger itself more orbital gravitational energy will be released.
In the first violent ignition channel there is no need to explain the additional release of gravitational energy, as it becomes part of the
energy budget of the explosion.
In the second channel of several years to tens of years from merger to explosion, the merger process will release energy.
In section \ref{sec:violent} we explain how this energy is dissipated such that the equatorial expanding gas keeps its relatively low velocity of $\sim 65-100 \km \s^{-1}$.
Basically, the released orbital energy is carried by a fast bipolar outflow and by radiation.
Here we note that the ejected mass will remove angular momentum as well.
Basically, the {{ disrupted}} object, which in our specific case   is the core of the giant, forms an accretion disk around the WD. {{ Since}}
the viscous timescale {{ of this disk}} is much shorter than a year {{ it}} plays no role in delaying the formation of a Chandrasekhar-mass remnant, that is formed within tens of years.
{{However, viscous dissipation}}
 removes angular momentum from the disk, {{ which must be carried away by the ejected mass. This can be done either by a disk wind or by jets, which}} carry
all the angular momentum and a substantial fraction of the excess energy (see section \ref{sec:violent}).

 {{ As previously discussed,}} most of the left-over mass from the merger process, which is a CO-rich gas, escapes in {{ the form of}} a disk wind or jets.
{{ Thus}}, the circumstellar gas will have a bipolar structure, similar to that of bipolar planetary nebulae.
{{ Since}} in the equatorial plane there will be the torus of the hydrogen-rich gas ejected from the envelope of the giant {{ moving slowly at velocities
$\sim 100 \km \s^{-1}$,  the  gas ejected by the explosion will first interact with it}}.
The CO-rich gas will be ejected at much higher velocities, {{ of the order of a few}} $\times 1000 \km \s^{-1}$ (the escape speed from the merger remnant),
and the encounter should take place much later, years to tens of years after explosion.
The collision with the gas influences the lightcurve (e.g., \citealt{Fryeretal2010}).

One argument against the WD-WD merger scenario for SN Ia is that there is an early off-center ignition of carbon.
However, most calculations of ignition in WD-WD merger were done for cold WDs (e.g., \citealt{Kawaietal1987}). Specifically, these authors found an early off-center carbon ignition but they consider a maximum mass accretion rate of $2 \times 10^{-5} ~{\rm M_{\sun}~yr^{-1}}$.
{{ In our scenario a much higher accretion rate is required}}. We actually have a process where the hot high-density core {{ of the giant}} is accreted {{ onto a more massive WD, for which there is a lack of self-consistent, realistic calculations}}.
Future simulations, {{ which are beyond the scope of this paper,}}  must check  whether {{ in this situation}}  excessive compression {{ can be avoided, and thus}}  prevent an early ignition of the WD.
Else, the ignition in the very dense region can set the explosion itself,
as in \cite{Pakmoretal2012b}. {{ All in all}}, new simulations are required to study the WD-core merger, using  sophisticated three-dimensional numerical codes, {{ because the outcome of the process is  very sensitive to numerical resolution (\citealt{Pakmoretal2012a}).}}

{{ It could be argued as well that} traditional SN Ia carbon ignition models require the detonation to be preceded by a deflagration  to occur over
$\sim 1000 \yr$, which is too long for the present scenario.
{{ However,}} we note that in recent years other ignition channels {{ have been}} discussed in the literature, e.g. ignition in
sub-Chandrasekhar WDs, and the violent merger ignition \citep{Pakmoretal2012b}.
{{ Consequently,} the ignition process must be reconsidered in the case of core-WD merger. For example, as the
accreted core gas is hot an ignition might occur in an intermediate regions where temperature is high
during the post-merger process.

\subsection{The violent merger phase}
\label{sec:violent}

When two WDs merge to form a super-Chandrasekhar mass WD a huge amount
of energy  is released.  Actually,  this might {{ result in}} an  event brighter
than expected from  SN Ia explosion if there is  a large delay between
merger and  explosion.  This is less of  a problem in the  case of the
violent merger scenario  \citep{Pakmoretal2012b} where the explosion occurs
several minutes after  merger {{ starts}}.  In the present  case, the merger
is of  a core of an  AGB star and a  WD companion (the  remnant of the
primary  star),  with  masses  of  $M_{\rm core}$  and  $M_{\rm  WD}$,
respectively.  {{ Nevertheless,  the merger}} of massive degenerate
components
releases less energy {{ as they reach}} $M_{\rm Ch}$.

The final  state just before the  SN Ia explosion is that the WD
accreted a mass $\Delta M$ from the  core, and the rest of the mass is
{{ ejected}}.
{{{ The {{ ejected}}  mass carries with it energy and angular momentum.  }}}
The difference  in binding  energy is  between  the energy
required to unbind  the core and the {{ gravitational}} energy  released by the {{ mass accreted}}
 onto the WD:
\begin{equation}
\Delta E_G \simeq 0.5 \frac {G (M_{\rm WD} + 0.5 \Delta M) \Delta M}{R_{\rm WD}} -
\Gamma \frac {G M_{\rm core}^2}{R_{\rm core}},
\label{eq:energy1}
\end{equation}
where  $\Gamma  \simeq  0.5$  is  a  parameter  that  depends  on  the
properties of the  core. We assume an adiabatic  index of $\gamma=5/3$
as the core is hot, and take  a coefficient of $0.5$ in the first term
of Eq.~(\ref{eq:energy1}).
{{{  The core and the WD first collide when they are on a highly eccentric orbit --- see section \ref{sec:migration}
and \cite{KashiSoker2011}. This implies that the kinetic energy of the binary system must be considered.
This energy is already included in the first term of Eq.~(\ref{eq:energy1}).   }}}
For  the typical parameters used here
the gravitational energy released is
\begin{eqnarray}
\Delta E_G ({\rm erg})&\simeq&
1.9 \times 10^{50}                                     
\left( \frac {M_{\rm WD} + 0.5 \Delta M}{1.3~{\rm M_{\sun}}}\right)
\left( \frac {\Delta M}{0.6~{\rm M_{\sun}}}\right)\nonumber\\
&\,& \left( \frac{R_{\rm WD}}{5400 \km} \right)^{-1}
-1.6\times 10^{50}                                     
\left(\frac{\Gamma}{0.6} \right)\nonumber\\
&\,& \left(\frac {M_{\rm core}}{1~{\rm M_{\sun}}}\right)^2
  \left( \frac{R_{\rm core}}{10^4 \km} \right)^{-1}.
\label{eq:energy2}
\end{eqnarray}

The radius of a  WD  of  mass   $1~{\rm M_{\sun}}$   is   $6000  \km$
\citep{Provencal1998}.   But  as  matter  is accreted  onto  the  cold
massive WD  its radius  shrinks. However, this  will not  release much
energy as the matter in  the center {{ of the WD}} is relativistic, and {{ thus}} the adiabatic
index $\gamma$  is close  to the full  relativistic value,  $4/3$.  For
$\gamma=4/3$, the  marginally-stable value,  no energy is  released as
the star contracts to satisfy hydrostatic equilibrium.  Here it  is not fully
relativistic, but  the adiabatic  index is still  {{ well}} below $5/3$  and not
much  energy will  be released.   The binding  energy of  a WD  at the
critical  mass  of $1.4~{\rm M_{\sun}}$  is  $E_b=-5.5\times 10^{50}  \erg$
(e.g., \citealt{IbanezCabanell1984}).  For a WD merger product of mass
$1.4~{\rm M_{\sun}}$ and  $\gamma=5/3$  the corresponding  radius is  $4700
\km$.  {{ In summary,}} we  conclude that  not  much energy  will  be  released by  the
contraction of the  cold accreting  WD, and took an effective radius of
$5400 \km$.

{{{ The ejected mass has more specific energy than it had before the merger. Therefore, it will carry some of the additional energy. As a large fraction of the ejected mass is blown
by an accretion disk wind, it will carry angular momentum and will {{ flow}} in the polar directions
(and might even form two opposite jets).    }}}

{{ Adopting typical values,}} the total energy that has  to be radiated {\it
before}  the explosion  is $E_{\rm  pre} \la  {\rm  several} \times
10^{49} \erg$.   This {{ estimate}} is very sensitive  to the parameters  due to the
substraction  of two very  close large  values. For  example, adopting
$M_{\rm WD} = 1.1~{\rm M_{\sun}}$ and  $\Delta M = 0.45~{\rm M_{\sun}}$, and using
the same  formalism we  find the released  energy to be  $\sim 10^{49}
\erg$.  The total energy released in radiation can be larger than that
emitted during  the SN explosion  itself. However, the  model assumes
(something that, {{ as mentioned earlier,}}   needs verification) that  the building of  the merger
remnant to  the point it explodes  lasts for several  years.  {{ Therefore,}} the timescale  is  two  orders  of  magnitudes  longer than  that  of  the  SN
explosion {{ itself}}.   The  bolometric luminosity  will  be  over  one order  of
magnitude lower than {{ that of}} the SN  explosion, and most of the radiation (due
to the optically  thick ejected envelope) {{ will}} be emitted  in the IR.  This
is  the  reason {{ why}} it  avoided  detection  by  different surveys  in  the
visible.   However, the  model predicts  that  such SN  Ia might  have
{{ a}} precursor  lasting  several  years  with 3--7  bolometric  magnitudes
fainter than the SN explosion, and  peaks in the near IR.  In case the
merger is  of a lighter  core-WD system, then the  pre-explosion total
energy released might be much larger.  But in that case the delay will
be longer, as  the central region needs to cool  to reach the critical
limit for explosion.

\section{THE EVOLUTIONARY ROUTE}
\label{sec:ignition}

\begin{figure}
\begin{center}
 \includegraphics[scale=0.5]{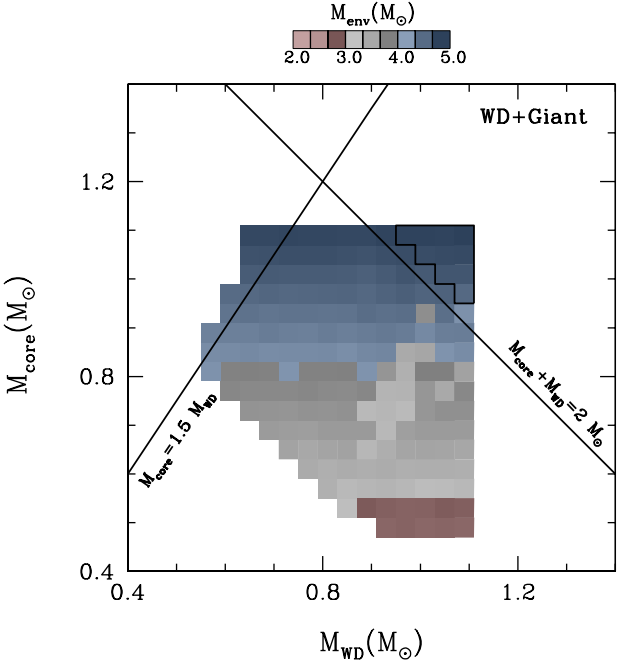}
        \caption{Mass of  the envelope of  the giant as a  function of
          the mass of  the primary white dwarf and of  the mass of the
          hydrogen-exhausted  core  of  the secondary.  The  different
          color scales  denote different masses of  the envelope.  The
          brown scale  corresponds to  relatively small masses  of the
          envelope ($2  \leq M_{\rm env}/ {\rm M_{\sun}} \leq  3$), the gray
          scale  corresponds to  intermediate masses  ($3  \leq M_{\rm
          env}/  {\rm M_{\sun}} \leq  4$),  while the  blue  scale has  been
          selected  to show the  masses of  interest for  the scenario
          described in  the main text  ($4 \leq M_{\rm  env}/ {\rm M_{\sun}}
          \leq  5$).   Also  plotted   are  the  constrains  given  by
          conditions 1  and 4.   For the sake  of clarity,  the region
          corresponding to the likely progenitors of PTF 11kx has been
          highlighted.  }
          \label{fig:massenv}
\end{center}
\end{figure}

For the  progenitor of  the explosion we  look for systems  having the
following properties.

\begin{figure}
\begin{center}
 \includegraphics[scale=0.5]{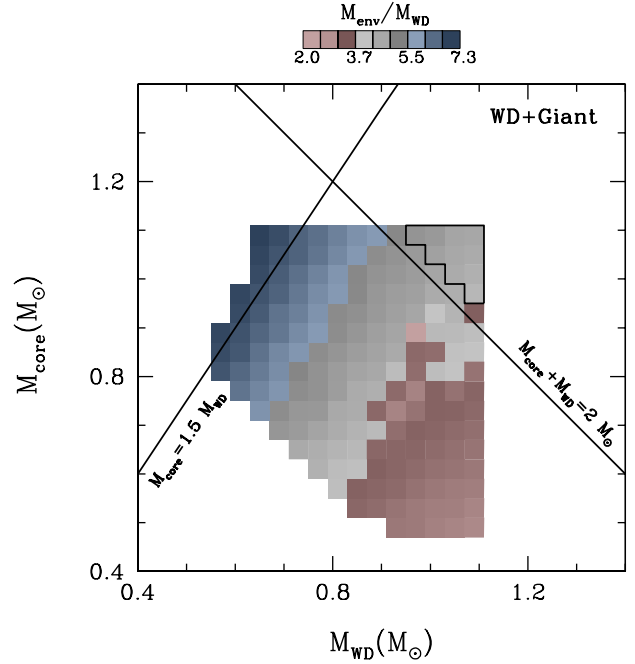}
        \caption{Ratio of the mass of the envelope of the secondary to
          the mass of the primary white dwarf, as given by the colored
          bar, as  a function of the  mass of the  primary white dwarf
          and of  the hydrogen-exhausted  core of the  secondary.  The
          rest    of   the    lines    are   the    same   shown    in
          figure~\ref{fig:massenv}.   Again,   the  region  of  likely
          progenitors of PTF 11kx has been conveniently highlighted in
          this plane. }
          \label{fig:massratio}
\end{center}
\end{figure}

\begin{enumerate}
\item The total  mass of the secondary core during  the final CE phase
  and  the mass  of  the WD  remnant  of the  primary  star should  be
  super-Chandrasekhar,  $M_{\rm core}  + M_{\rm  WD} >  1.4~{\rm M_{\sun}}$.
  But  as discussed  in  the  previous section,  to  prevent a  pre-SN
  explosion outburst,  the core  and WD should  be much  more massive,
  $M_{\rm core} + M_{\rm WD} \ga 2~{\rm M_{\sun}}$.
\item The total mass of the envelope  at the final CE phase must be as
  massive  as the  observed CSM.  We take  it to  be $M_{\rm  env} \ga
  2~{\rm M_{\sun}}$, but prefer $M_{\rm env} \ga 4~{\rm M_{\sun}}$.
\item  To facilitate  merger at  the termination  of the  CE  phase we
  require $M_{\rm env}/M_{\rm  WD} \ga 3$, but prefer  a larger number
  even \citep{Soker2013}.
\item It is  likely that an explosion without a  delay will take place
  when the core is accreted onto  the cooler WD remnant of the primary
  star \citep{KashiSoker2011}. For that the density of the core should
  be lower than the density of the  WD.  As the core is hot its radius
  is larger than that of a  cool WD. The condition reads $M_{\rm core}
  \la 1.5~M_{\rm WD}$.
\item As  we require the merger  product to be  a CO WD, to  avoid ONe
  WDs,  we are  limited  to both  $M_{\rm  core} <  1.1~{\rm M_{\sun}}$  and
  $M_{\rm WD}<1.1~{\rm M_{\sun}}$.
\end{enumerate}

To  examine for  such systems  we  ran the  population synthesis  code
described  in \cite{Garcia-Berro2012},  where all  details  are given.
Here, for  the sake  of conciseness, we  only present the  results for
systems that obey the  constraints above.  In figure \ref{fig:massenv}
we plot  the envelope mass of the  secondary at the time  of merger in
the   plane   $M_{\rm  core}$ -- $M_{\rm   WD}$,   while  in   figure
\ref{fig:massratio} we present  the ratio of the envelope  mass to the
WD mass.  Note that the  likely progenitors of  PTF 11kx are  within a
region of WD masses between  $\sim 0.9$ and $\sim 1.1~{\rm M_{\sun}}$, while
the masses of  the hydrogen-exhausted core are also  within $\sim 0.9$
and  $\sim  1.1~{\rm M_{\sun}}$.   The  most  restrictive  conditions  are,
respectively,  conditions   1  and  3.   Namely,  we   find  that  the
progenitors of this supernova must  fulfill that the total mass of the
merged remnant should  be larger than $2~{\rm M_{\sun}}$,  and also that the
ratio of  the envelope mass to the  mass of the white  dwarf should be
larger than $\sim 3$, although other possibilities cannot be presently
discarded.

\cite{Dilday2012} crudely  estimate that the  fraction of SNe  Ia that
exhibit  prominent  circumstellar interaction  near  maximum light  is
$\sim 0.1 -1 \%$.  As $\sim 1$--$2$ SN Ia occur per $1000 {\rm M_{\sun}}$ stars
formed \citep{Maoz2012},  the estimate of  \cite{Dilday2012} stands at
$\sim 0.001  -0.02$ SN Ia with  massive CSM per  $1000~{\rm M_{\sun}}$ stars
formed.  Our Monte Carlo simulations  show that the frequency of these
systems  with  the  conditions  listed  above  is  $0.002$  per  $1000~{\rm M_{\sun}}$ stars  formed.  If  we relax the  assumption on  the core+WD
mass  and the  envelope mass  to $M_{\rm  core} +  M_{\rm WD}  \ga 1.8~{\rm M_{\sun}}$ (instead of $M_{\rm core} + M_{\rm WD} \ga 2~{\rm M_{\sun}}$), and
$M_{\rm   env}  \ga   0.5~{\rm M_{\sun}}$  (instead   of  $M_{\rm   env}  \ga
4~{\rm M_{\sun}}$),  respectively,  we  find  $0.016$  systems  to  obey  the
condition per  $1000~{\rm M_{\sun}}$ stars formed.  The  later envelope mass
might be appropriate  to SN~2005gj.  We conclude that  the {{{{  {violent-prompt merger } }}}} scenario
can  satisfactorily explain  all the  observed properties  of PTF~11kx
and, moreover,  that the number  of expected progenitors  with massive
CSM is consistent with the observations.

\section{SUMMARY}
\label{sec:summary}

The conclusion that the presence of any CSM around a SN Ia implies its
association  with  the   single-degenerate  (SD)  scenario  ---  e.g.,
\cite{Sternberg2011} --- is problematic,  in particular in cases where
the  CSM mass  within  $\sim 1000  \AU$  is $\ga  0.01~{\rm M_{\sun}}$.
An explosion set  by the {{{{  {violent-prompt} }}}} merger  of a WD  companion with the core  of the
giant  star naturally  occurs within  a  massive CSM  --- the  ejected
common envelope \citep{Livio2003}.


We find the association of  the SN Ia PTF~11kx \citep{Dilday2012} with
the SD  scenario to be unlikely  due to the massive  CSM.  Instead, we
found that the {{{{  {violent-prompt} }}}}  merger of a massive WD with a  massive core, as marked
by    the   thick    line    on   the    upper    right   corner    of
Figures~\ref{fig:massenv}  and  \ref{fig:massratio},  to  be  able  to
account for  the properties of PTF~11kx.  We  predict that interaction
of the ejecta with the CSM will take place over the coming decades and
that the amount of mass in the CSM will be accumulated to few times of
${\rm M_{\sun}}$.  We also found  from our population synthesis calculations
that the  number of systems with large  total mass of the  WD and core
match  the  number  of  SN   Ia  with  massive  CSM  as  deduced  from
observations.


\section*{Acknowledgments}
This research  was supported by the  Asher Fund for  Space Research at
the  Technion, by MCINN  grant AYA2011--23102,  by the  European Union
FEDER funds, and by the ESF EUROGENESIS project (grant EUI2009-04167).


\label{lastpage}

\end{document}